\documentstyle[epsfig,mncite]{mn}
\ifCUPmtlplainloaded \else

\fi
\title[Keck II Spectroscopy of mHz QPOs in Hercules X-1]{Keck II spectroscopy of mHz Quasi-Periodic Oscillations in Hercules X-1}

\author[K. O'Brien et al.]{K. O'Brien$^{1,2}$, Keith Horne$^1$, B.
Boroson$^{3,}\thanks{National Research Council Associate}$, M.
Still$^{3,}\thanks{Universities Space Research Association}$, R.
Gomer$^4$, J. B. Oke$^{5,6}$, \cr P. Boyd$^3$, S. D. Vrtilek$^7$\\
        $^1$ School of Physics and Astronomy, University of St. Andrews, St. Andrews KY16 9SS \\
        $^2$ Astronomical Institute ``Anton Pannekoek'', University of Amsterdam, 1098-SJ Amsterdam, The Netherlands \\
	$^3$ NASA/Goddard Space Flight Center, Code 662, Greenbelt, MD 20771, USA \\
        $^4$ Howard Hughes Medical Institute and Department of Biochemistry and
Cell Biology, MS-140, Rice University, Houston, \\ TX 77005-1892, USA \\
        $^5$ California Institute of Technology, Mail Stop 105-24, Pasadena, CA 91125, USA\\ 
	$^6$ Dominion Astrophysical Observatory, Herzberg Institute of Astrophysics, National Research Council of Canada, \\ 5071 West Saanich Road, Victoria, BC V8X 4M6, Canada \\
	$^7$ Harvard-Smithsonian Center for Astrophysics, 60 Garden Street, Cambridge, MA 02138, USA \\}

\date{Accepted ... . Received ...; in original form 2000 ....}

\pagerange{\pageref{firstpage}--\pageref{lastpage}}

\begin{document}
%
%
%
\newcommand{\novasco}{GRO\,J1655-40}
\newcommand{\sco}{Scorpius~X-1}
\newcommand{\her}{Hercules~X-1}
\newcommand{\hzher}{HZ Herculis}
\newcommand{\cyg}{Cygnus~X-2}
%
%
\newcommand{\HST} {{\it HST}}
\newcommand{\xte} {{\it RXTE}}
\newcommand{\rxte}{{\it RXTE}}   
\newcommand{\XTE} {{\it RXTE}}
\newcommand{\RXTE}{{\it RXTE}}
\newcommand{\GRO} {{\it GRO}}
%
%
\newcommand{\HI} {H\,\textsc{i}}
\newcommand{\HII} {H\,\textsc{ii}}
\newcommand{\HeI} {He\,\textsc{i}}  
\newcommand{\HeII} {He\,\textsc{ii}}
\newcommand{\HeIII}{He\,\textsc{iii}}
%
%
\newcommand{\EBV}{E(B-V)}
\newcommand{\Rv} {R_{\rm V}}
\newcommand{\Av} {A_{\rm V}}
%
%
\newcommand{\lam} {$\lambda$}
\newcommand{\lamlam}{$\lambda\lambda$}
%
%
\newcommand{\comm}[1]{\textit{[#1]}}
\newcommand{\etal}{{\it et al}.\ }
\newcommand{\mdot}{\dot{M}}
\newcommand{\msolar}{M_{\sun}}
\newcommand{\msun}{M_{\sun}}
\newcommand{\degree}{$^{\circ}$}
\newcommand{\rank}{S$_z$}
%
%
\newcommand{\object}{Hercules X-1}
\newcommand{\utstart}{07:40}
\newcommand{\utend}{08:50}
\newcommand{\etime}{72.075}
\newcommand{\skyfrac}{0.9}
\newcommand{\nspec}{58,000}
\newcommand{\binfac}{7}
\newcommand{\bintime}{0.5045}
\newcommand{\nbin}{9484}
%

\label{firstpage}

\maketitle
\begin{abstract}

We present Keck II spectroscopy of an optical mHz quasi-periodic oscillations (QPOs) in the lightcurve of the X-ray pulsar binary \her. In the power spectrum it appears as `peaked noise', with a coherency $\sim$2, a central frequency of 35 mHz and a peak-to-peak amplitude of 5\%. However, the dynamic power spectrum shows it to be an intermittent QPO, with a lifetime of $\sim$hundred seconds, as expected if the lifetime of the orbiting material is equal to the thermal timescale of the inner disk. We have decomposed the spectral time series into constant and variable components and used blackbody fits to the resulting spectra to characterise the spectrum of the QPO variability and constrain possible production sites. We find that the spectrum of the QPO is best-fit by a small hot region, possibly the inner regions of the accretion disk, where the ballistic accretion stream impacts onto the disk. The lack of any excess power around the QPO frequency in the X-ray power spectrum, created using simultaneous lightcurves from \rxte, implies that the QPO is not simply reprocessed X-ray variability.

\end{abstract}


\section{Introduction}

\her/\hzher\ belongs to the sub-class of LMXBs known as X-ray
pulsar binaries. \her\ is a 1.4 $\msolar$ neutron star with a 1.24~s spin
period. It is in a 1.7 day binary orbit with it's $\sim 2 \msolar$
companion star, \hzher. Due to the high inclination of the system, $i
\simeq 85$\degree, it shows deep eclipses on the orbital period. 

The variability of \her\ has been very well studied at all wavelengths. In
the X-rays the spin of the neutron star is observed, modulated by the
orbital eclipses and a third 35-day cycle. This 35-day X-ray periodicity
was first discovered by \scite{tananbaum72} using the $\it{UHURU}$
satellite as on X-ray on-state followed by an extended off-state.
\scite{jones76} later reported a weaker short-on state during the
off-state. The 35-day periodicity consists of a main-on state, lasting
about 10 days, and a short-on state lasting 5 days, separated by periods
where the X-ray flux drops to almost zero \cite{scott99}. The X-ray
pulsations are
clearly seen in the X-rays during the on-states but appear with a much
lower amplitude during the off states, as expected if the cycle is caused
by the obscuration of the central regions of the accretion disk. The
evolution of the pulse profile between and during these on-states has been
well studied but the origins and evolution of it's complex structure
remain poorly defined.

While the existence and form of the 35-day cycle is well documented, the
mechanism responsible is less well understood. It is thought to be caused
by a warped, precessing accretion disk that occults the neutron star at
certain disk phases. \scite{pringle96} showed that a strongly irradiated
accretion disk is unstable to warping. The radiation pressure acting on
the inner regions of the accretion disk forms a warped precessing
accretion disk as seen in \her \cite{maloney97,wijers99}. One other model is that
the warp in the disk can be formed by an X-ray driven coronal wind
\cite{schandl94}. In this model, the X-rays from the central compact object
irradiate the inner regions of the accretion disk and produce a hot
corona, when the sound velocity of the gas exceeds the escape velocity of
the potential well the material leaves the binary as a coronal wind,
exerting a repulsive force on the disk. \scite{schandl94} showed that,
with suitable boundary conditions, they could create a suitable warp to
explain the general features of the 35-day cycle. This model can also
explain the observed anomalous and pre-eclipse dips as an interaction of
the accretion stream with the warped disk, causing disk thickness
increases that temporarily obscure the line of sight to the X-ray source,
causing a dip in the X-ray flux \cite{schandl96}.

At optical wavelengths, much of the light is due to X-ray heating of the
inner face of the companion star and other regions of the binary. This
effect is seen to modulate on the binary phase and with the 35-day
super-period, implying that the companion star is irradiated even during
the observed X-ray off states. The most notable evidence for reprocessing
of X-rays on the companion star of \her\ comes from the observed optical
pulsations \cite{middleditch76}.

We have found a series of low frequency optical Quasi-Periodic Oscillation (QPO) with a central frequency of 35 mHz. These QPOs have also been observed in the UV \cite{boroson2000} with a frequency of 45 mHz. While these observations
are not simultaneous with the optical data presented here, the similarity
of the frequencies implies a similar production mechanism. QPOs have been
observed in many binary systems, most noticeably in the X-ray observations
of both persistent and transient X-ray binaries. In Z-sources, such as
\sco\ and \cyg, the QPOs are seen to change in both frequency and power as
the mass transfer rate onto the compact object changes \cite{hasinger89}. 
In transient sources similar X-ray QPOs are seen. The transient X-ray
pulsar, XTE~J1858+034 shows 110 mHz QPOs with an RMS amplitude of 3.7-7.8
\%. \scite{chakrabarty98} found both X-ray and optical QPOs in the X-ray
pulsar 4U 1626-67, with a frequency of 48 mHz. The optical QPO in this
system is most likely due to reprocessing off the innerface of the
strongly irradiated companion star of an X-ray QPO. This X-ray QPO is
possibly caused by 'blobs' on Keplerian orbits outside the corotation
radius of the neutron star \cite{kommers98}. 

In this paper we describe the optical characteristics of the QPO, in
order to constrain the position of the region in the binary producing the
variability and infer possible production mechanisms. 

%
%
%
%
\section{Observations}
\begin{table}
\begin{center}
\begin{tabular}{|c|c|c|}
\hline
Parameter & Start & End \\ \hline
UT & 07:40 & 08:50 \\
MJD & 51000.32018 & 51000.36879 \\
Binary phase & 0.407 & 0.435\\
35-day phase & 0.524 & 0.526 \\ \hline
\end{tabular}
\end{center}
\caption{Summary of optical observations. The binary phase is calculated using the ephemeris of \protect\scite{deeter91} and the 35$-$day phase uses the ephemeris from \protect\cite{still2000}.}
\label{obstable}
\end{table}
The optical data was taken on 1998 July 6, using the Low Resolution Imaging Spectrograph
(LRIS; \scite{oke95}) on the 10-m Keck II telescope on Mauna Kea, Hawaii. A summary of the observations is given in Table~\ref{obstable}. The LRIS was used with a 5.2
arcsecond slit masked with aluminized mylar tape to form a square
aperture.  The 300/5000 grating used has a mean dispersion of 2.55
\AA/pixel in the range 3600\AA\ - 9200\AA. We used a novel data
acquisition
system to obtain more than \nspec\ spectra of \her, in the
form of a continuous byte stream. The mean integration
time was measured to be \etime\-ms and there is no dead-time between individual
spectra. In addition to the 2048-pixel spectrum, 25-pixel under-scan and 75-pixel over-scan regions were used to measure the CCD bias level, which we subtracted from each spectrum. The noise for a given pixel was
calculated using a readout noise of 6.3 e$^-$ and a gain of 4.7 e$^-$/ADU. Cosmic
rays were rejected with a threshold of 10-sigma from the de-biased frames.
A master flatfield image was created by finding the median of 700
individual flatfield spectra. 
 This image showed no deviations above 0.3 \% in all but 3 pixels. It was
decided that it was therefore not necessary to flatfield individual
spectra. Calibration arc spectra and spectra of the sky in the region of
the object were taken at regular intervals. 

The background sky spectrum, which accounts for $\sim$ \skyfrac\% of the total flux was subtracted using the best fit sky spectrum. Sky spectra were taken at the beginning and end of the run, so that long timescale variations could be detected. The mean and variable components of the sky spectrum were found by creating a lightcurve for each pixel and extracting the mean and gradient of this lightcurve. These coefficients were then filtered in wavelength, with a running median filter of width 101 pixels. The gradient of the data was found to be almost zero, as the sky brightness didn't change noticeably during the run. The arc calibration was done by fitting a second order polynomial to 7 lines in a median spectra of exposures of Hg and Ar lamps. Arc spectra from the beginning and end of the run were used to take into account any drifts in the wavelength scale. The wavelength calibration was applied using the MOLLY spectral analysis package. The individual spectra were flux calibrated using exposures taken of the standard star, Feige 56 \cite{oke90}. A low order polynomial fit was found to the median of all the individual flux star spectra and this calibration was again applied using MOLLY. 

%
%
\subsection{Time calibration}
Due to the nature of the continuous stream of data, it was impossible to attach accurate timemarks to each individual spectrum. Individual time marks, accurate to approximately 200-ms, were placed after every other spectrum using the computer clock. In order to establish an absolute time reference, secondary timestamps were obtained on many occasions during the 5-night observing run. To create these timestamps, an incandescent lamp that illuminated the CCD was modulated in synch with radio broadcast time signals. From these we found an ephemeris and mean exposure time. These were checked against the computer's time marks and found to be in excellent agreement. 

In order to increase the signal to noise of the individual spectra and make the data-set more manageable, \binfac\ individual spectra were binned together to give \nbin\ binned spectra with a time resolution of \bintime\ seconds.

%
%
\subsection{X-ray observations}

The observations took place on 1998 July 6, using the PCA onboard \rxte.
Lightcurves were created from the {\sc E\_250us\_128M\_0\_1s} mode data-set. The energy
range of the PCA is 2-100\,keV. The lightcurves were corrected to the
solar-system barycentre, using the subroutine {\sc fxbary} in the
{\sc ftools} software suite. They were also corrected for the motion of
the neutron star in its binary orbit around its 2.2$\msolar$ companion
\hzher, using the orbital ephemeris of \scite{still2000}.

%
%
\section{Results}
The continuum lightcurve created from our optical dataset, in the
wavelength range 5000\AA\ - 5800\AA, is shown in the upper panel of
Figure~\ref{totallc}. A subset of this lightcurve, clearly showing the QPO
is shown in the lower panel of Figure~\ref{totallc}.

We used two methods, one temporal and one spectral, to characterise the variability of the source. We created power spectra of the source to determine the temporal characteristics of the variability, then we used this information to determine the spectral characteristics by means of a variability spectrum.

\subsection{The power spectrum}
\begin{figure}
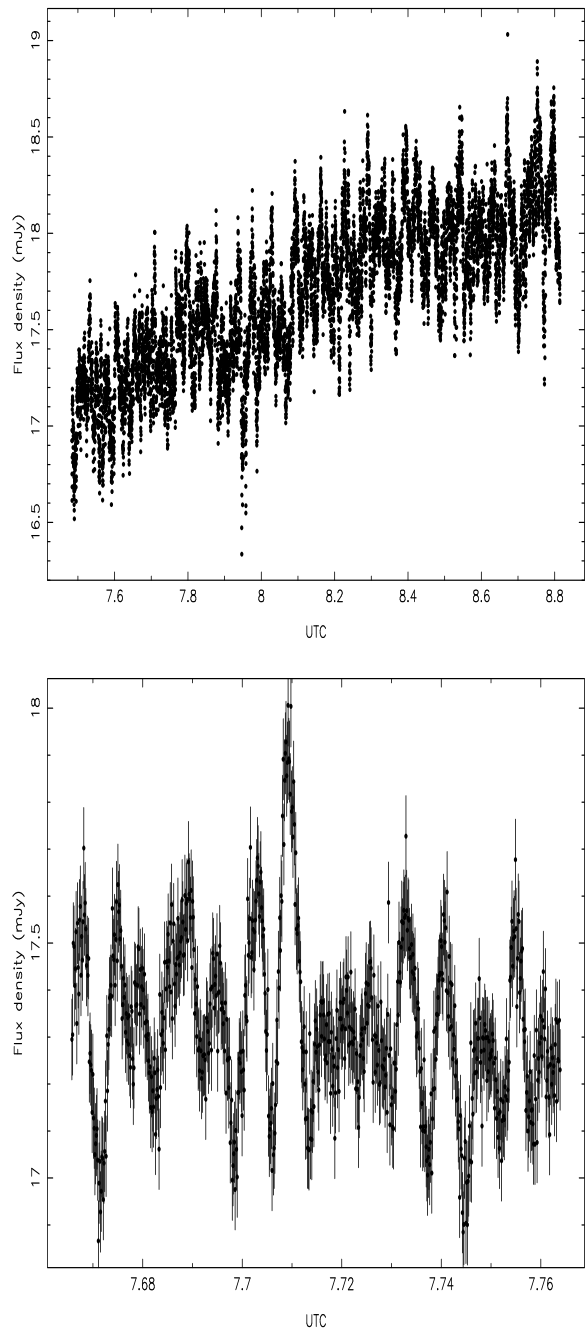

\begin{center}
\epsfig{width=3.3in,height=3in,angle=270,file=fig1a.ps}

\vspace{5mm}

\epsfig{width=3.4in,height=3in,angle=270,file=fig1b.ps}

\caption{Top panel, lightcurve of the optical variability in \her\ for the
entire data-set. Bottom panel, lightcurve of a subset of the data clearly
showing the QPO. The units of flux density are mJy in the wavelength
range 5000 - 5800 $\AA$. Each data point is the average of approximately
7 individual spectra. No error bars are shown in the top panel for
clarity.}
\label{totallc}
\end{center}
\end{figure}

\begin{figure*}
\begin{center}

\epsfig{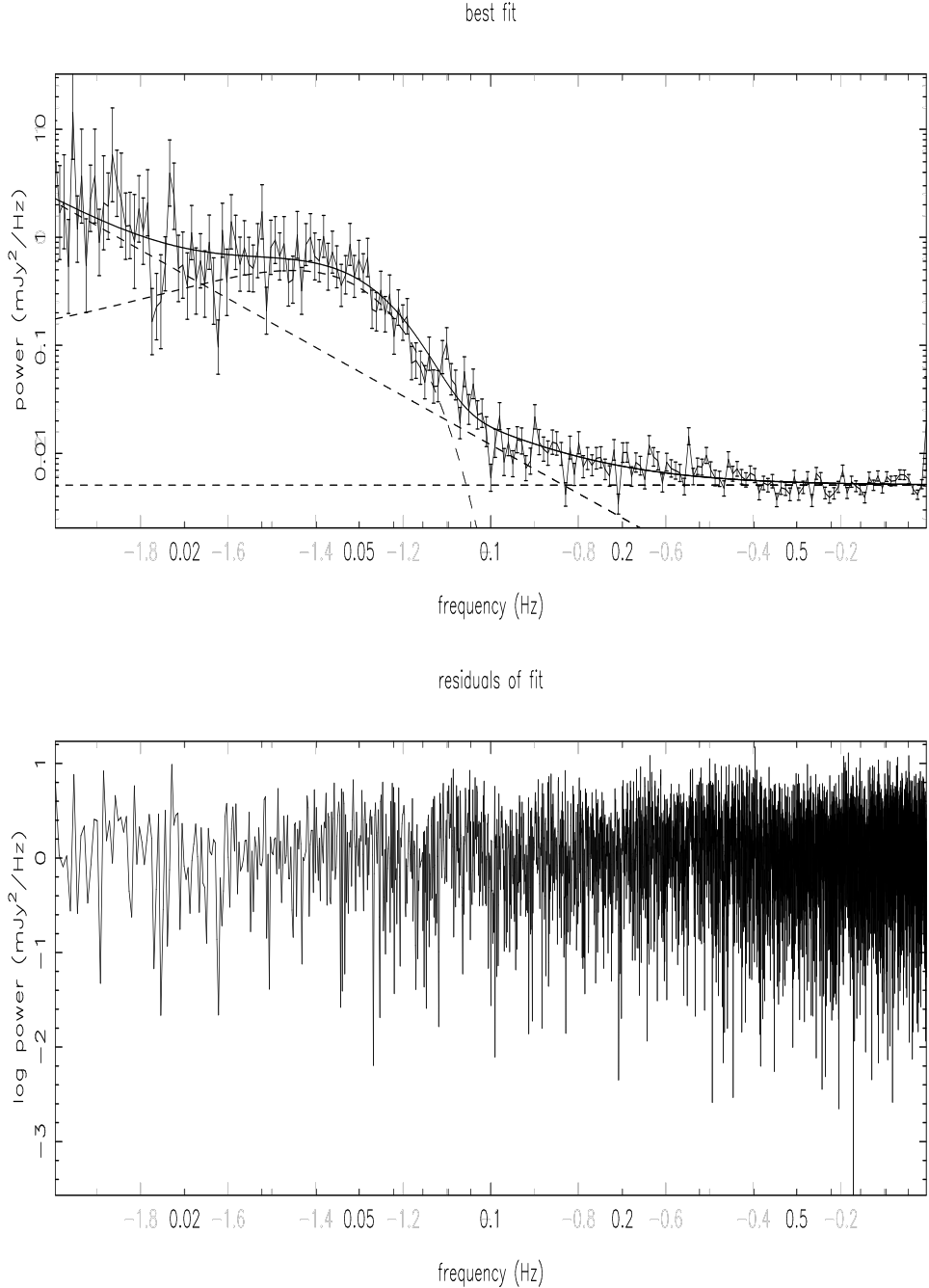}

\caption{Top, the binned power spectrum for the total dataset, showing the best fit model (solid line), with a reduced $\chi^2 = 1.581$ for 4687 degrees of freedom. The white noise, power law and QPO components (dashed lines) are also shown based on their best fit parameters. Bottom, the residuals of the fit, showing the full resolution of the power spectrum.}
\label{pspec1}
\end{center}
\end{figure*}

We created power spectra for the total data-set by simple sine-curve fitting to the input lightcurve, in a method analogous to the Lomb-Scargle periodogram \cite{press89}. However in our method, optimal weights were applied when fitting each sine-curve. This method was preferred to fast Fourier transforms (FFTs) due to the presence of irregular gaps in the data. The resulting power spectrum is shown in Figure~\ref{pspec1}. The power spectrum was represented as a sum of three terms; (1) white noise power at all frequencies, (2) a `red noise' power-law describing the intrinsic flickering of the source, and (3) a hump-like peaked noise feature, caused by the summation of the individual transient QPOs, described as a Gaussian. These features can be described numerically as,
\begin{eqnarray}
P\left(f\right) & = &P_{wn}\left(f\right) + P_{pl}\left(f\right) + P_{qpo}\left(f\right) \nonumber \\
& = & P_{wn} \nonumber \\
& + & P_{pl} \left(\frac{f}{f_{p}}\right)^{-\alpha} \nonumber \\
& + & P_{qpo} \exp\left[{- \frac{1}{2} \left(\frac{f - f_{0}}{\Delta f}\right)^{2}}\right] ,
\label{pspeceqn}
\end{eqnarray}
where $P_{wn}$ is the power of the white noise level, $P_{pl}$ is the power at $f_{p}$, fixed to be 0.1 Hz, of the `red noise' power-law with spectral index $\alpha$. $P_{qpo}$ is the peak power of the Gaussian QPO feature with a central frequency $f_0$ and standard deviation $\Delta f$. (Note: while a lorentzian is normally used to characterize a QPO, we have reverted to using a Gaussian to characterize the randomness of the position of the central frequencies of the individual QPOs)

The coherence if the QPO is low,
\begin{equation}
Q = \frac{f_{0}}{\Delta f} \sim 2,
\end{equation}
suggesting that the sinusoidal components change period, phase or amplitude substantially in just a few cycles, as can be seen in the bottom panel of Figure~\ref{totallc}.

The best fit values of the free parameters are shown in
Table~\ref{bestvals}. These were found using an amoeba code to search
through the parameter space, see \scite{press92} for a discussion of
amoeba codes. The best fit has a $\chi^{2}$ of $7410.15$ for $4687$
degrees of freedom. We determined the 1-parameter 1-sigma confidence
regions for each parameter by fixing a parameter and then optimising the
fit until the badness of fit is equal to $\chi^{2}_{aim}$.
$\chi^{2}_{aim}$ is determined using the value of $\chi^{2}_{min}$, where,
\begin{equation}
\chi^{2}_{aim} = \chi^{2}_{min} + \frac{\chi^{2}_{min}}{N-P},
\label{chieqn}
\end{equation}
where $N-P$ is the number of degrees of freedom in the fit.
\begin{table}
\begin{center}
\begin{tabular}{|c|c|c|}
\hline
Parameter & unit & value \\ \hline
$P_{wn}$ & $10^{-3}\,mJy^{2}\,Hz^{-1}$ & $5.09^{+0.14}_{-0.15}$\\
$\alpha$ & & $2.25^{+0.13}_{-0.18}$\\
$P_{pl}$ & $10^{-3}\,mJy^{2}\,Hz^{-1}$ & $12.0^{+1.4}_{-1.4}$\\
$P_{qpo}$ & $10^{-3}\,mJy^{2}\,Hz^{-1}$ & $49.30^{+10.4}_{-7.1}$\\
$f_{0}$ & mHz & $35.2^{+4.3}_{-8.1}$\\
$\Delta f$ & mHz & $17.4^{+3.7}_{-2.6}$\\ 
Integrated QPO & $10^{-2}\,mJy^{2}$ & $21.5^{+6.4}_{-4.5}$\\ 
Power & & \\
%
\end{tabular}
\end{center}
\caption{Best-fit values to model for power spectrum of total dataset, shown in Equation~\ref{pspeceqn}. The best fit $\chi^2$ was $7410.15$ for $4687$ degrees of freedom.}
\label{bestvals}
\end{table}
\subsection{Dynamic power spectrum}
We used the Gabor transform (Heil \& Walnut 1989,1990)\nocite{heil89}\nocite{heil90} to create dynamic power spectra to investigate the time evolution of the power spectrum determined from our optical data. This is especially valuable due to the low coherency of the oscillations, to determine if the low coherency is caused by the evolution of the central frequency of the oscillation, perhaps caused by material rapidly moving through the inner regions of the accretion disk, or whether the low coherency is caused by randomly occurring events based around a central frequency, as would be the case for short-lived blobs of material on Keplerian orbits within the disk.

\begin{figure*}
\begin{center}
\epsfig{width=6.6in,height=6.6in,angle=270,file=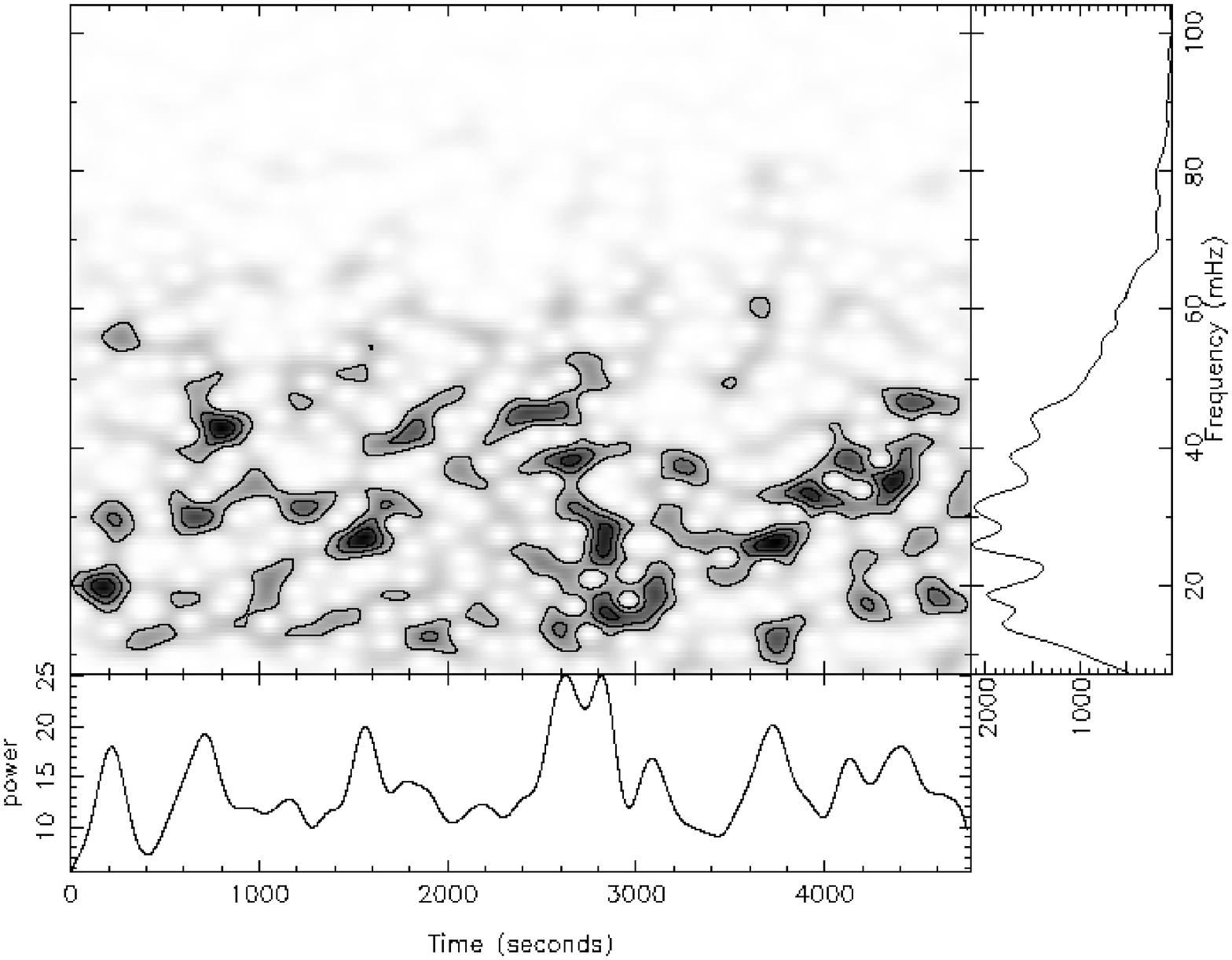}

\caption{Results from performing a Gabor transform on the filtered optical
dataset. The main panel shows the dynamic power spectrum. The right-hand
panel shows the average power spectrum and the lower panel shows the
integrated power as a function of time.}

\label{gaborplot}
\end{center}
\end{figure*}

The discrete Gabor transform is a wavelet transform that allows us to
determine the power spectrum over a sub-set of the data, using a Gaussian
window on the lightcurve \cite{boyd95}. We have chosen the 3-sigma width
of the window to be 750-s. The data set contained 9484 data points,
which have been filtered with a running median filter with a filter width
of 101-s to remove the slowly varying component. For computational 
purposes, the resulting
lightcurve was padded with 6900 randomly distributed points with the same
mean and variance as the data set. The results of this analysis are shown
in Figure~\ref{gaborplot}, where the main image shows the results of the
Gabor transform, the lower panel shows the integrated power at a given
time and the right-hand panel shows the
 average power spectrum. 

The dynamic power spectrum shows transient QPOs in the frequency range of the peaked noise feature. While there is little evidence of frequency evolution of the individual features, it is possible that the feature that appears at $\sim$ 15 mHz at T~=~2800-s, evolves into the feature at $\sim$ 46 mHz at T~=~4500-s. From the coherency of the individual QPOS, we see that the average lifetime of the features is 200-300-s, or ten times the Kepler period of the features. 

\subsection{A reprocessed X-ray QPO\,?}

In order to determine if the optical QPO is a reprocessed version of an
X-ray QPO, we calculated power spectra of the X-ray data simultaneous
with our optical data. We chopped the total X-ray lightcurve into 512-s sections and created individual Leahy-normalised power spectra \cite{leahy83} for
each lightcurve. We then combined the separate power spectra to
create a total power spectrum for the observations. This method is more
successful at determining the power in the QPO, which spans a broad range of frequencies and is only coherent on short timescales. 

\begin{figure}
\begin{center}
\epsfig{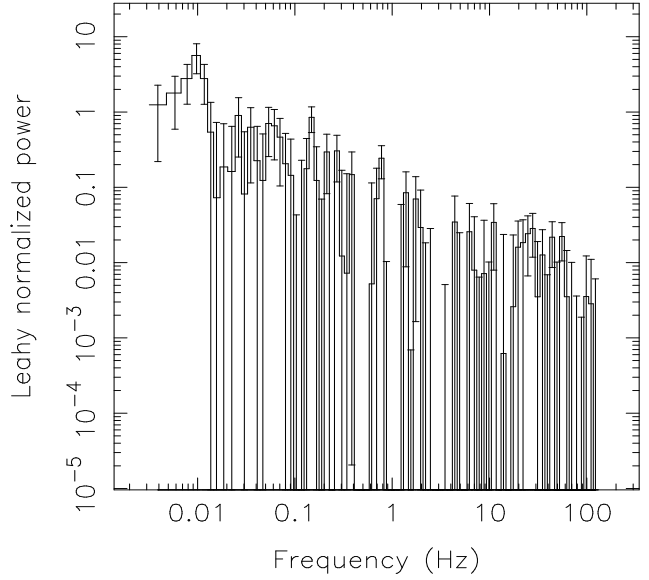}
\caption{The X-ray power spectrum for our RXTE observations, simultaneous with the optical observations. The power spectrum has been normalised, following the method of {\protect \scite{leahy83}}, and the Poisson level has been removed. The power spectrum shows no evidence for excess power in the region of the optical QPO. However an excess of power is clearly seen at 0.01 Hz.}
\label{x1pspecplot}
\end{center}
\end{figure}

The average power spectrum, shown in Figure~\ref{x1pspecplot}, shows no evidence for the QPO at 35 mHz, implying the optical QPO is not simply an X-ray QPO that is reprocessed into an optical flux by some region in the binary. However there is evidence for some excess power at 0.01 Hz, which is possibly the X-ray counterpart of our slow variability.

\subsection{The spectrum of the variability}

We created variability spectrum-lightcurve pairs that, when re-combined, reproduce the entire spectral time series, using a method similar to that used by \scite{eracleous96} to decompose the flaring and quiescent spectra of AE Aqr. The model of the spectral time series, $M(\lambda,t)$, is written as, 
\begin{equation}
M\left(\lambda,t\right) = S_{0} \left(\lambda\right) + S_{1}\left(\lambda\right)L_{1}\left(t\right) + S_{2}\left(\lambda\right)L_{2}\left(t\right),
\end{equation}
where $S_{0}\left(\lambda\right)$ is the average spectrum, $S_{1}\left(\lambda\right)$ and $L_{1}\left(t\right)$, are the spectrum and lightcurve of the first variable component and $S_{2}\left(\lambda\right)$ and $L_{2}\left(t\right)$ are the spectrum and lightcurve of the second variable component. $L_{1}\left(t\right)$ and $L_{2}\left(t\right)$ are normalised to a mean value of zero and unit variance. This normalisation ensures that the spectral components $S_{1}\left(\lambda\right)$ and $S_{2}\left(\lambda\right)$ have units of mJy and do not contribute to the time averaged spectrum. 

The model is compared to the observed spectral time series and the badness of fit calculated using a $\chi^2$ analysis. We solved iteratively for each component of the model, until a suitable convergence criterion was met, namely that the percentage change in the global $\chi^2$ was less than 0.1\% over 3 successive iterations. In order to check that a global minimum had been reached, we chose several different sets of initial values, each of which converged onto the same final values. 

In order to separate the variability into slow and fast components we filtered the lightcurves, using Savitzky-Golay filters \cite{press92} of different widths. We low-pass filtered the lightcurve $L_1(t)$ to remove variations faster than 80-s. This component then measures primarily the red noise power-law component identified in the power spectrum. Similarly we low-pass filtered the lightcurve $L_2(t)$ to remove variations faster than 5-s, which are primarily the white noise component. This means that $L_2(t)$ responds to variations between 80-s and 5-s, which include the tail of the red noise and the QPO. 

\begin{figure*}
\begin{center}
\epsfig{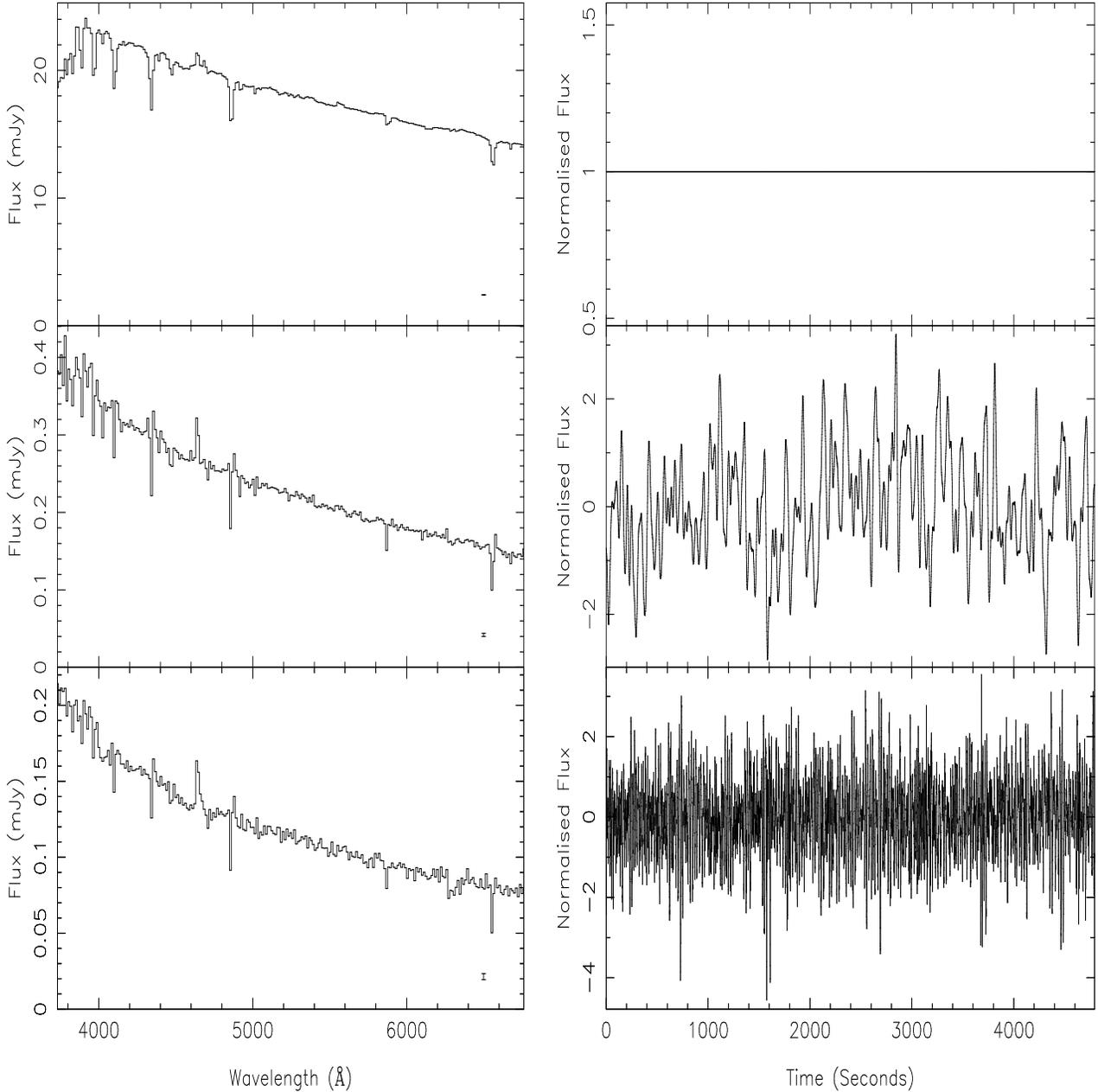}

\caption{Top panels, the average spectrum, $S_{0}\left(\lambda\right)$,  (left) and its associated
lightcurve (right), which is set to be unity. Centre panels, the first
variability spectrum, $S_{1}\left(\lambda\right)$, (left) and its associated lightcurve, $L_{1}\left(t\right)$, (right). Bottom
panels, the same as above for the second component of the variability. Typical error bars are shown on the spectra. The
first variability lightcurve is detrended and filtered with a width 40-s. The second variability lightcurve is filtered with a width of 5-s.}

\label{hervarspec1}
\label{hervarspec2}
\end{center}
\end{figure*}

The results of the the fit to the spectral time series are shown in Figure~\ref{hervarspec1}. In the first variability lightcurve, $L_{1}\left(t\right)$ the fluctuations seen with frequencies around 10 mHz are resolved. The spectrum of this variability, $S_{1}\left(\lambda\right)$, is bluer than the average spectrum, $S_{0}\left(\lambda\right)$. The rms amplitude of this variable component increases from 1\% of the mean flux at the red end of the spectrum to 2\% at the blue end. The second variability lightcurve, $L_{2}\left(t\right)$, shows the variability of the 35 mHz QPO. The features are transient, showing the QPO is not coherent over more than a few cycles. The spectrum of this variable component, $S_{2}\left(\lambda\right)$, has an rms amplitude of 0.5 \% at the red end of the spectrum to 1 \% in the blue. The spectral features in the two variability spectra are very similar. The Balmer lines appear to be narrower than in the average spectrum, implying they have variable wings, while the cores of the lines vary less than the average spectrum. The NII + CII blend at 4641 \AA\ also appears to show a strong variable component. This line is interesting as it is a blend of two lines formed by Bowen fluorescence \cite{schacter89} and is thought to be caused by irradiation of the binary by the central X-ray source. In contrast the He II line at 4685 \AA\ is not apparent in either variability spectrum, implying that it is not highly variable.

These results agree well with the power spectra analysis in the previous section. The average flux in the second variability lightcurve, shown in the third panel in Figure~\ref{hervarspec2}, is in good agreement with the flux in the QPO and the variability lightcurve shows variability on the same timescales as the QPO period. We conclude from this that the second variability spectrum, $S_{2}\left(\lambda\right)$, represents the spectrum of the QPO.

\subsection{Spectral fits to variability spectra}
The next stage of the analysis was to fit model spectra to the data to fit physical parameters to the observed average and variability spectra from the previous analysis. To do this we used two models to see which best represents the data. Both models use blackbody spectra to describe the observed continuum fluxes. 
\begin{enumerate}
\item
A single blackbody:

This will tell us the temperature and size of the emitting region. The model spectrum is defined at a wavelength $\lambda$ as,
\begin{equation}
F_{\nu}(\lambda) = \frac{\pi R^2}{D^2}\, B_{\nu}(\lambda,T) 
\end{equation}
for a region with a radius,$R$, a temperature, $T$, at a distance $D$. This model is most useful for a region that has a roughly constant temperature for the duration of the observations.

\item
The difference of two blackbodies:

The second model is fitting the difference of two blackbodies to the data.
This involves fitting the difference of two blackbodies with different temperatures, $T_{1}$ and $T_{2}$, where $T_{2} > T_{1}$. The model spectrum is defined at a wavelength $\lambda$ as,
\begin{equation}
F_{\nu}(\lambda) = \frac{\pi R^2}{D^2}\, \left[ B_{\nu}(\lambda,T_{2}) - B_{\nu}(\lambda,T_{1}) \right]
\end{equation}
for a region with a radius, $R$, at a distance $D$. This model is most useful for describing a region whose temperature varies during the observations. These variations are limited to coming from the same region, ie a single region where the temperature varies between $T_{1}$ and $T_{2}$.
\end{enumerate}

\begin{figure*}
\begin{center}
\epsfig{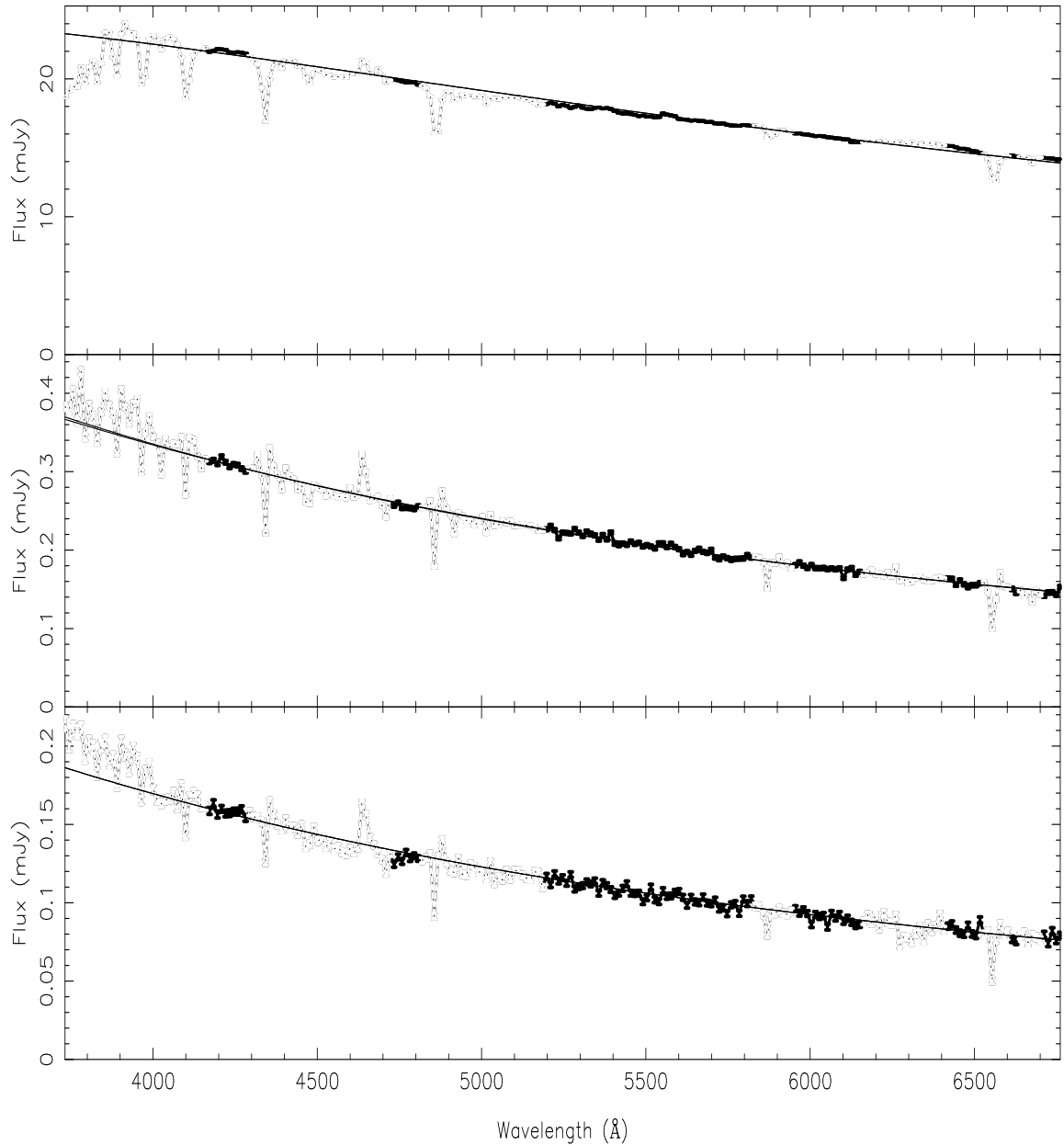}

\caption{The average spectrum (top panel) and the spectra of the slowly varying and QPO components (middle and bottom panels respectively). The dotted lines show the data points not used in the fitting routine, the solid points are the data used and the solid line is the best fit model.}
\label{varfitplot1}

\end{center}
\end{figure*}

These models were fitted to regions of the spectra that are least contaminated with lines. The best fit values, together with their 1-parameter 1-sigma confidence regions are shown in Table~\ref{varfittable}. The data together with the best fit models and the regions used are shown in Figure~\ref{varfitplot1}. The value of $\chi^2$ is large in all three cases, due to the high signal-to-noise of each spectrum revealing intrinsic deviations from a Blackbody. However, with the confidence intervals re-scaled as shown in Equation~\ref{chieqn}, the parameters remain reasonably well constrained.

\begin{table*}
\begin{center}
\begin{tabular}{|c|c|c|c|c|}
\hline
& unit & average spectrum & slow variability & fast variability \\
& & $S_{0}$ & $S_{1}$ & $S_{2}$ \\ \hline
model 1: & & & & \\
Temp & K & 16500 $^{+100}_{-100}$ & 37000 $^{+1000}_{-1000}$ & 33600 $^{+1500}_{-1400}$ \\
Radius & $10^{10}$ cm & 29.36 $^{+0.19}_{-0.20}$ & 1.64 $^{+0.04}_{-0.03}$ & 1.26 $^{+0.03}_{-0.04}$ \\
$\chi^{2}/100$ & & 9016 & 3.919 & 2.966 \\ \hline 
model 2: & & & & \\
$T_{1}$ & K & 600 $^{+1900}_{-\inf}$ & 7600 $^{+9900}_{-3800}$ & 600 $^{+4400}_{-\inf}$\\
$T_{2}$ & K & 16500 $^{+100}_{-100}$ & 31400 $^{+5700}_{-14600}$& 33600 $^{+1500}_{-1700}$\\
Radius & $10^{10}$ cm & 29.4 $^{+0.2}_{-0.2}$ & 1.89 $^{+\inf}_{-0.24}$ & 1.26 $^{+0.04}_{-0.04}$ \\
$\chi^2/99$ & & 9107 & 3.904 & 2.996 \\ \hline 
\end{tabular}
\end{center}
\caption{Summary of best fit parameters to the spectral fitting of the spectra from the variability model, together with their 1-parameter 1-sigma confidence regions. A distance of 5 kpc was assumed throughout.}
\label{varfittable}
\end{table*}
%

%
%
\section{Discussion}

We have studied the X-ray and optical variability in the accreting pulsar \her\ and found evidence for a series of optical QPOs that comprise a broad peaked noise feature in the overall power spectrum. While the coherence of this feature is low, $Q \sim 2$, the individual QPOs that comprise the peak have much higher coherencies, as seen in the lightcurve in the lower panel of Figure~\ref{totallc} and the widths of the features in the dynamic power spectrum. We have characterised this noise feature as having a Gaussian shape in the power spectrum. This has been fitted along with the white noise level and a power law to describe the variability of \her\ at low frequencies. The best fit to this model has a peak frequency of 35.2 mHz. 

We have extracted the spectrum of the variability from the frequency range of the QPO. We have also determined the average spectrum and the spectrum of any slower variability, that has characteristic frequencies from the red noise, power-law region of the power spectrum, as shown in Figure~\ref{hervarspec1}. The variability spectra are bluer and fainter than the average spectrum, indicating they are from small, high temperature regions. The average spectrum is predominantly that of the companion star, \hzher. It shows strong absorption in the Balmer lines, as well as some He I absorption lines. There is some evidence of the irradiation of the companion in the form of a weak He II ($\lambda$4686) emission line and Bowen fluorescence lines ($\lambda$4634, $\lambda$4641, $\lambda$4642), which can be used as a diagnostic of the physical conditions in the binary \cite{schacter89}. The variability spectra both show a bluer continuum than the average spectrum with Balmer and Helium I absorption lines, however in these spectra the wings of the lines show more variability than the cores. The Bowen lines also show more variability than the average spectrum, as expected if the variability is coming from a region that has been heated by the irradiation of the X-ray source.

Our simple blackbody fits to the observed spectra, while not particularly physical, enable us to constrain models for their production. The best fit model to the average spectrum is a single temperature blackbody, with a mean temperature of 16500 K. The size of the emitting region is $8.6 \times 10^{22}$ cm$^2$. While this temperature is inconsistent with the temperature of the companion star in the absence of irradiation, which has been observed during eclipse of the accretion disk to have a temperature of $\sim$ 7000 K, it is most likely to be due to the strongly irradiated inner face of the companion star. The size of emitting region would cover approximately 70\% of the innerface of the companion star, assuming a distance of 5 kpc and an effective Roche lobe radius of $3.5 \times 10^{11}$ cm \cite{eggleton83}. 

The best fit model to the variability spectra of the slowly varying component also enable us to place constraints on the size and possible locations of the emitting region. The slowly varying component is best fit with the derivative of a blackbody model, although the single temperature blackbody is almost as good a fit. In this model the source varies between 7600 K and 31000 K, from a region of $3.6 \times 10^{20}$ cm$^2$. One candidate for this component is the same region on the companion star as the average spectrum, but with a smaller area, possibly the most highly irradiated regions close to the inner Lagrangian point. 

The best fit model to the spectrum of the QPO is that of a single temperature blackbody with a mean temperature of $\sim$ 33600 K, with an area of $1.6 \times 10^{20}$ cm$^2$. The data is not consistent with the model of a derivative of a blackbody as used to describe the slowly varying component. The fit converges to almost the same solution as with the single temperature Blackbody. This leads us to consider whether or not the QPO is produced in a physically distinct region from that of the slowly varying component. The QPO producing region is a small, high temperature region, probably in the accretion disk surrounding the compact object. Therefore if the frequency of the QPO is the Kepler frequency of material orbiting around the neutron star, the Kepler radius of the material orbiting with a frequency of 35 mHz around a 1.3 $\msolar$ neutron star would be $3.3 \times 10^{9}$ cm. This would appear to be too small to contain the region responsible for the observed spectrum. However if we take the {\it range} of frequencies included in the QPO, we find that the Kepler radii covered are $ 9.4 \times 10^{8}$ - $5 \times 10^{10}$ cm, which is large enough to encompass the observed region. 

The existence of a range of preferred azimuths in the disk explains the observations of the QPO frequencies. There are two mechanisms whereby this anisotropy could cause the observed QPO. The first is the material in the preferred region of the disk is orbiting the neutron star and we are seeing the irradiated innerface of the region, as it passes across the backside of the disk. The second explanation is that the azimuths are fixed in the plane of the binary and that inhomogeneities in the material traveling through these regions gives rise to enhanced emission. This material would orbit the neutron star and emit when it passes through the preferred azimuths. 

\scite{schandl96} modelled the warp in the accretion disk to explain the occurance of the anomalous and pre-eclipse dips, using a coronal wind model. In this model it is the variation with synodic period of the penetration depth of the stream into the inner regions of the accretion that accounts for an increase in material along the line of sight to the binary that accounts for the observed dips. The Keck II observations were taken at the synodic orbital phase of 0.894. At this phase the accretion stream can penetrate down to the inner regions of the accretion disk. According to Schandl's calculations at this synodic period the stream will penetrate the disk at a radius of $\sim$ $5.1 \times 10^{10}$ cm. This is in excellent agreement with the lowest Kepler frequency of the observed QPO. This model predicts that the thickness of the disk will increase as the stream interacts with the disk. If this energy is dissipated on a thermal timescale, then for a thin disk, $t_{dyn} \sim \alpha\,t_{th}$, so that $\alpha \sim Q^{-1}$. Thus our observation that the coherence of an individual feature in the dynamic power spectrum last for $\sim 10 \times t_{dyn}$ naturally leads to a value of $\alpha \sim 0.1$.

Such a model, where the disk-stream interaction point leads to a blob of material orbiting in the disk, close to the corotation radius, could also explain the observations of the QPO in 4U 1626-67. In this system, the QPO has a similarly low coherence, but is observable over more than a decade \cite{kommers98}, which is easily explained if the accretion stream causes a relatively stable preferred radius in the disk. However in \her\ we do not observe the QPO in the X-ray lightcurves, as in 4U 1626-67, although this could be due to the viewing angle during the X-ray off state. Observations during the X-ray on state are necessary, with simultaneous optical observations to determine if the optical QPO is indeed a reprocessed X-ray QPO.

\section*{acknowledgments}
KOB was supported by a PPARC research Studentship for much of this work. We thank John Cromer for writing, testing, and loading the software that allowed the LRIS CCD to read out continuously, and Bob Leach for helpful discussions.  We especially thank Tom Bida and Frederic Chaffee for kindly letting us make changes to the LRIS system. Data presented herein were obtained at the W.M. Keck Observatory, which is operated as a scientific partnership among the California Institute of Technology, the University of California and the National Aeronautics and Space Administration.  The Observatory was made possible by the generous financial support of the W.M. Keck Foundation. We would also like to thank Peter Jonker for his help with the X-ray power spectrum analysis. This research has made use of NASA's Astrophysics Data System Abstract Service.

\bibliographystyle{mn}
\bibliography{/home/kso/Papers/ksobib}

\label{lastpage}

\end{document}